\documentclass{svproc}
\usepackage{type1cm}        
\usepackage{graphicx}        

\usepackage[bottom]{footmisc}

\usepackage{newtxmath}       

\usepackage{xcolor}
\usepackage[colorlinks=true, allcolors=blue]
{hyperref}

\usepackage{url}

\usepackage{booktabs,parskip}    
\usepackage{multirow}   
\usepackage{multicol}

\usepackage{caption}
\usepackage{subcaption}

\definecolor{umass}{rgb}{0.004, 0.3215, 0.5568}
\definecolor{etonblue}{rgb}{0.59, 0.78, 0.64}
 \definecolor{mycolor}{rgb}{0.12, 0.3, 0.17}
\definecolor{myblue}{rgb}{0.03, 0.27, 0.49}
\definecolor{mediumjunglegreen}{rgb}{0.11, 0.21, 0.18}
\definecolor{asparagus}{rgb}{0.53, 0.66, 0.42}
\definecolor{goldenpoppy}{rgb}{0.99, 0.76, 0.0}
\definecolor{lincolngreen}{rgb}{0.11, 0.35, 0.02}
\definecolor{red(ncs)}{rgb}{0.77, 0.01, 0.2}
\definecolor{darkorange}{rgb}{1.0, 0.55, 0.0}
\definecolor{safetyorange(blazeorange)}{rgb}{1.0, 0.4, 0.0}
\definecolor{smokeytopaz}{rgb}{0.58, 0.25, 0.03}
\definecolor{brightgreen}{rgb}{0.4, 1.0, 0.0}
\definecolor{tue}{rgb}{0.82, 0.14, 0.14}

\begin{document}
	\mainmatter              
	\title{Multi-way contingency tables with uniform margins}
	\titlerunning{Multi-way contingency tables with uniform margins}  
	%
	\author{Elisa Perrone\inst{1} \and Roberto Fontana\inst{2}
	 \and Fabio Rapallo\inst{3}}
	\authorrunning{Elisa Perrone et al.} 
	%
	\tocauthor{Elisa Perrone, Roberto Fontana, Fabio Rapallo}
	\institute{Department of Mathematics and Computer Science, Eindhoven University of Technology, Groene Loper 3, 5612 AE Eindhoven, The Netherlands, \email{e.perrone@tue.nl}
		\and
		Department of Mathematical Sciences, Politecnico di Torino, Corso Duca degli Abruzzi 24, 10129 Torino, Italy, \\
    \email{roberto.fontana@polito.it}
        \and
        Dipartimento DIEC, Universit\`{a} di Genova, via Vivaldi 5, 16126 Genova, Italy,    \email{fabio.rapallo@unige.it}}
	\maketitle              
	\begin{abstract}
We study the problem of transforming a multi-way contingency table into an equivalent table with uniform margins and same dependence structure. 
Such a problem relates to recent developments in copula modeling for discrete random vectors. 
Here, we focus on three-way binary tables and show that, even in such a simple case, the situation is quite different than for two-way tables. Many more constraints are needed to ensure a unique solution to the problem. Therefore, the uniqueness of the transformed table is subject to arbitrary choices of the practitioner. We illustrate the theory through some examples, and conclude with a discussion on the topic and future research directions.

	\end{abstract}
	\section{Preliminaries}
    \label{sec:1}
Tabular data in the form of contingency tables appear in many applications, such as health care, biology, and social science. 
Such data format have extensively been investigated in statistics with the primary goal of developing tools to extract information about the relations between the variables (see \cite{rudas2018lectures}).
In \cite{geenens2020copula}, the author draws interesting connections between the analysis of bivariate contingency tables and \emph{copulas}. 
We recall that the key concept behind copula theory is the separation between the marginal effect and the dependence structure. Such a separation allows for \emph{ad hoc} modeling tools for dependence and is possible by transforming the original joint probability distribution into one with uniform margins. The transformed distribution is the copula associated with the original distribution and is uniquely defined if the margins are continuous (\cite{sklar_59}). The question if we can benefit from a similar idea in the discrete context has been investigated in \cite{geenens2020copula}.
There and in the references therein, the authors explore the concept of transforming a given two-way contingency table into a new one with uniform margins. Such a transformation makes it easier to interpret the underlying association of the table which might be hidden when margins are heavily unbalanced, as discussed below. 

\begin{figure}
\centering
\begin{subfigure}{.5\textwidth}
  \centering
  \includegraphics[width=1\linewidth]{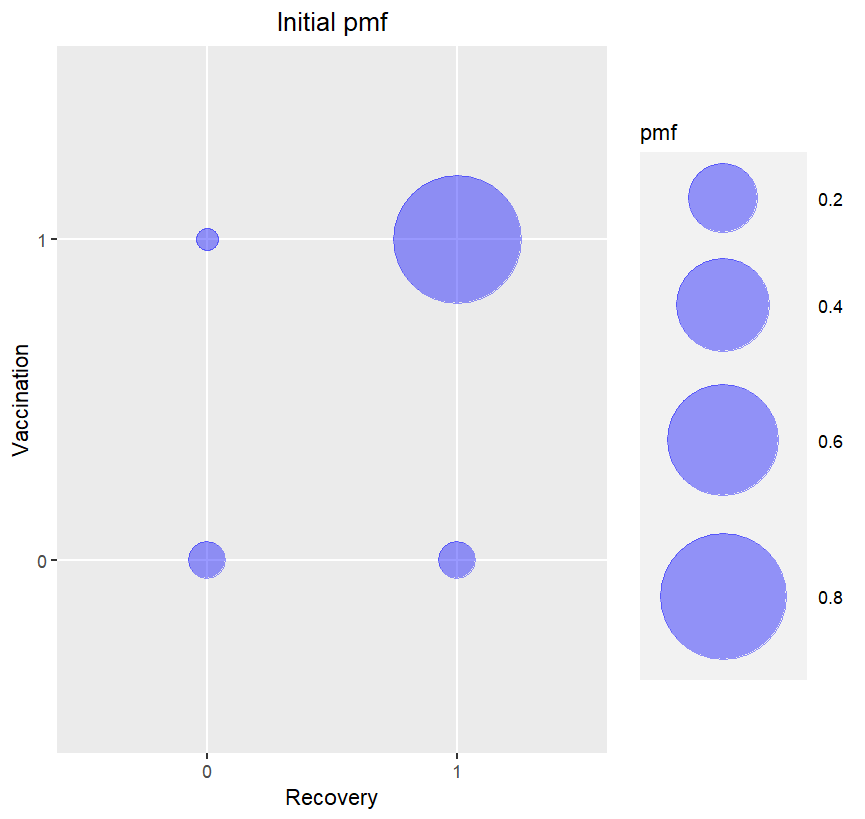}
  \caption{Original data}
  \label{fig:sub1}
\end{subfigure}%
\begin{subfigure}{.5\textwidth}
  \centering
  \includegraphics[width=1\linewidth]{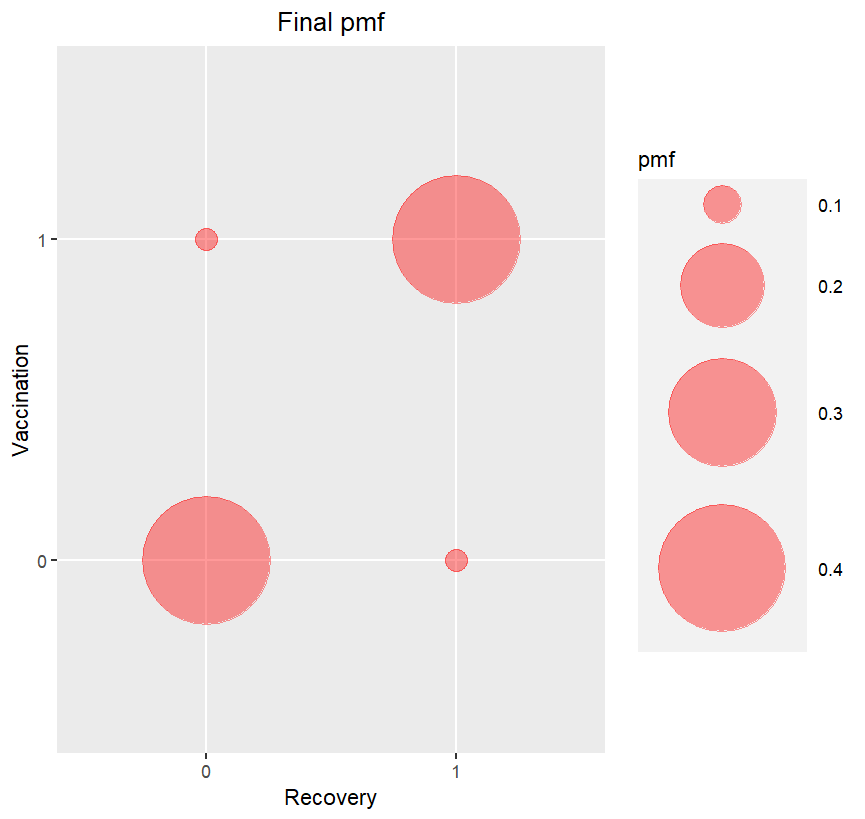}
  \caption{Transformed data}
  \label{fig:sub2}
\end{subfigure}
\caption{Bubble plot of original data (left) and transformed data (right)}
\label{fig:sheffield}
\end{figure}

For $2 \times 2$ contingency tables, the transformation entails converting the original table into a new one where all marginal probabilities equal $1/2$. Table \ref{tab:id7} reports a classical example from \cite{yule1912methods} where the data represents smallpox patients at Sheffield Hospital classified based on vaccination status [yes/no] and recovery [yes/no]. 
The odds ratio is substantial as it equals $19.47$. However, due to marginals that are far from uniform, the association may not be readily apparent in the original data. Nevertheless, the association becomes evident when the data is transformed to have marginal probabilities of $1/2$ while maintaining the same odds ratio of $19.47$. Figure \ref{fig:sheffield} shows the effect of the transformation on the visualization of the data.

In this work, we investigate how to extend the idea of identifying an equivalent contingency table with uniform margins when we have dimension greater than two. In particular, we focus on the simplest case of $2 \times 2 \times 2$ contingency tables. 
In Sect.~\ref{sec:2}, we introduce the notation and show our main results. Some interesting examples are illustrated in Sect.~\ref{sec:3}, and we conclude with open questions for future research on the topic in Sect.~\ref{sec:4}.
\begin{table}[h]
    \centering
\normalsize
    \begin{tabular}{cc|c|c}
 \hline
    Vaccination  ($X_1$) & Recovery ($X_2$)& \quad $\Tilde{n}$ \quad  & \quad  $\Tilde{p}$ \quad \\
    \hline
no & no & 274 & 0.06\\
no & yes & 278 & 0.06\\
yes & no & 200 & 0.04\\
yes & yes & 3951 & 0.84\\
\hline \\
    \end{tabular}
    \caption{Sheffield smallpox epidemic reported in \cite{yule1912methods}.}
    \label{tab:id7}
\end{table}

\section{Multi-way contingency tables and odds ratios}
\label{sec:2}
We here outline the mathematical framework introduced by Geenens in \cite{geenens2020copula} relevant for our work. We summarize the findings in two dimensions before moving the discussion to three dimensions.

\subsubsection{Two-way binary tables.}
Given a $2\times 2$ contingency table $(\Tilde{n}_{ij}, i,j=0,1)$ let $N=\Tilde{n}_{00}+\Tilde{n}_{11}+\Tilde{n}_{01}+\Tilde{n}_{10}$ be the grand total and   $\omega_\star=\frac{\Tilde{n}_{00}\Tilde{n}_{11}}{\Tilde{n}_{01}\Tilde{n}_{10}}$ the odds ratio. For our goal, we consider the table written in terms of the relative frequencies $\Tilde{p}_{ij}=\frac{\Tilde{n}_{ij}}{N}$ instead of the counts $\Tilde{n}_{ij}$. 
As explained in the motivating example of Sect.~\ref{sec:1}, the goal is to find a new table $(p_{ij} \geq 0, i,j=0,1)$ with marginals equal to $1/2$ that maintains the same odds ratio $\omega_\star$ of the original table. Such a table can be obtained by solving the system of equations in Eq.~(\ref{eq:d2})

\begin{equation}
\left\{
\begin{aligned}
\frac{p_{00} p_{11}}{p_{01} p_{10}}&=\omega_\star \\
p_{00} +p_{01} +p_{10} +p_{11} &=1\\
p_{00} +p_{01} - p_{10}  - p_{11} &=0\\
p_{00} - p_{01} +p_{10} - p_{11} &=0
\end{aligned}
\right. ,
\label{eq:d2}
\end{equation}
whose unique solution only depends on the odds ratio and is as follows
\begin{equation}
\left\{
\begin{aligned}
p_{00}=p_{11}=\frac{\sqrt{\omega_\star}}{2(1+\sqrt{\omega_\star})} \\
p_{01}=p_{10}=\frac{1}{2(1+\sqrt{\omega_\star})}
\end{aligned}
\right. .
\label{eq:sol2}
\end{equation}
Using the data in Table \ref{tab:id7}, we obtain $p_{00}=p_{11} \approx 0.41$ and $p_{01}=p_{10} \approx 0.09$.

\subsubsection{Three-way binary tables.}
We now consider three classification variables $X_1,X_2,X_3$, with relative frequencies denoted by $p_{i,j,k}$, for $i,j,k \in \{0,1\}$.
A possible generalization of the approach presented for two-way tables to three-way tables goes through the definition of the \emph{3-dimensional odds ratio} $\omega_3$ given as
\begin{equation}
\omega_3=\frac{p_{000} p_{011} p_{101} p_{110}}{p_{001} p_{010} p_{100} p_{111}} . 
\label{eq:omega3}
\end{equation}
Additionally, one can also define the so-called \emph{conditional odds ratios}:
\begin{equation}
\omega_2(X_i, X_j|X_k=a), \; i\neq j \neq k
\label{eq:condodd3}
\end{equation}
where $\omega_2(X_i, X_j|X_k=a)$ is the 2-dimensional odds ratio computed for $X_i$ and $X_j$ referring to the sub-table defined by $X_k=a$, $a \in \{0,1\}$. Since $k$ is uniquely determined by $i$ and $j$, we streamline the notation by writing $\omega_{ij|a}$ instead of $\omega_2(X_i, X_j|X_k=a)$.
Interestingly, the 3-dimensional odds ratio and the conditional odds ratios are linked by the following chain of equalities (see \cite{rudas2018lectures}):
\begin{equation}
\omega_3=\frac{\omega_{12|0}}{\omega_{12|1}} =\frac{\omega_{23|0}}{\omega_{23|1}} =\frac{\omega_{13|0}}{\omega_{13|1}}   .
\label{eq:prop3d}
\end{equation}
We are now ready to propose a possible generalization of the problem presented for two-way binary tables.
In particular, we consider a $2\times 2 \times 2$ contingency table $(\Tilde{n}_{ijk}, i,j,k=0,1)$, with grand total $N$, relative frequencies $\Tilde{p}_{ijk}=\Tilde{n}_{ijk}/N, i,j,k\in \{0,1\}$, 3-dimensional odds ratio $\omega_\star$ (defined as in Eq.~(\ref{eq:omega3})), and conditional odds ratios $\omega_{ij|a}^\star$ (defined as in Eq.~(\ref{eq:condodd3})). 
The goal is to find the relative frequencies of a new table $(p_{ijk} \geq 0, i,j,k \in \{0,1\})$ with uniform marginals equal to $1/2$ that maintains the same odds ratio $\omega_\star$ as the original table. Basically, we transform the problem of Eq.~(\ref{eq:d2}) into the system of equations reported below:
\begin{equation}
  \left\{
\begin{aligned}
\frac{p_{000} p_{011} p_{101} p_{110}}{p_{001} p_{010} p_{100} p_{111}} &=\omega_\star \\
p_{000} +p_{001} +p_{010} +p_{011} + p_{100} +p_{101} +p_{110} +p_{111} &=1\\
p_{000} +p_{001} +p_{010} +p_{011} - p_{100} - p_{101} - p_{110} - p_{111} &= 0\\
p_{000} +p_{001} -p_{010} -p_{011} + p_{100} + p_{101} - p_{110} - p_{111} &= 0\\
p_{000} - p_{001} +p_{010} -p_{011} + p_{100} - p_{101} + p_{110} - p_{111} &= 0\\
\end{aligned}
\right.  .
\label{eq:d3}
\end{equation}
Differently from the case of two-way tables, the solution of the system in Eq.~(\ref{eq:d3}) is not unique. One possible solution is given by the generalization of the 2-dimensional solution Eq.~(\ref{eq:sol2}) as follows
\begin{equation}
\mathbf{p}_\star=
\left\{
\begin{aligned}
p_{000}= p_{011}=p_{101}=p_{110}=\frac{\sqrt[4]{\omega_\star}}{4(1+\sqrt[4]{\omega_\star})} \\
p_{000}= p_{011}=p_{101}=p_{110}=\frac{1}{4(1+\sqrt[4]{\omega_\star})} 
\end{aligned}
\right. .
\label{eq:sol3}
\end{equation}
However, there are in general many more solutions suggesting that this straightforward extension is not enough to guarantee a prescribed dependence structure in the new table.
A way to ensure a unique solution is to add more constraints to the problem. 
In particular, a natural choice would be to require that the transformed table $p_{ijk}$ preserves also the conditional odds ratios of the original table. A new system of equations that includes the additional constraints is given in Eq.~(\ref{eq:omega3cond}). We notice that the relationships highlighted in Eq.~(\ref{eq:prop3d}) allows us to only use the conditional odds ratios  $\omega_{23|0}^\star, \omega_{13|0}^\star$, and $\omega_{12|0}^\star$, since the others can be derived by combining them with the 3-dimensional odds ratio.
The problem presented in Eq.~(\ref{eq:omega3cond}) has a unique solution that can be computed using symbolic mathematical software like Mathematica 
\begin{equation}
\small
\left\{
\begin{aligned}
\frac{p_{000} p_{011} p_{101} p_{110}}{p_{001} p_{010} p_{100} p_{111}} &=\omega_\star \\
\frac{p_{000} p_{011} }{p_{001} p_{010}} &=\omega_{23|0}^\star\\
{\frac{p_{000} p_{101} }{p_{001} p_{100}}} &{=\omega_{13|0}^\star}\\
{\frac{p_{000} p_{110} }{p_{010} p_{100}}} &{=\omega_{12|0}^\star}\\
p_{000} +p_{001} +p_{010} +p_{011} + p_{100} +p_{101} +p_{110} +p_{111} &=1\\
p_{000} +p_{001} +p_{010} +p_{011} - p_{100} - p_{101} - p_{110} - p_{111} &= 0\\
p_{000} +p_{001} -p_{010} -p_{011} + p_{100} + p_{101} - p_{110} - p_{111} &= 0\\
p_{000} - p_{001} +p_{010} -p_{011} + p_{100} - p_{101} + p_{110} - p_{111} &= 0\\
\end{aligned}
\right. .
\label{eq:omega3cond}
\end{equation}

\section{Examples}
\label{sec:3}

In this section, we present two examples to illustrate the results reported in Sect.~\ref{sec:2}. First, we analyze the artificial data of Table~\ref{tab:Agresti2}, taken from \cite{agresti:12}. 
The cross-classifications involves three binary variables, which are the response to a treatment (success or failure, $X_1$), the treatment (drug A or drug B, $X_2$), and the clinic (1 or 2, $X_3$). 
The frequencies in Table~\ref{tab:Agresti2} are expected counts from a model with conditional odds ratio $\omega_{12|0}^\star$ (treatment and response given the clinic) equal to 1. 
The other conditional odds-ratios are given by $\omega_{23|0}^\star= 6$, $\omega_{13|0}^\star=0.167$, and the 3-dimensional odds-ratio $\omega_\star=1$.
Through Mathematica, it was possible to find the solution $p$ to the problem of Eq.~(\ref{eq:omega3cond}). The solution is reported in Table~\ref{tab:Agresti2}. 
Again using Mathematica, we were also able to find the closed form solution to the more general problem of Eq.~(\ref{eq:d3}) with the only constraint (besides uniform margins) that the 3-dimensional odds ratio equals 1. The solution has three free variables and is reported in the following Eq.~(\ref{eq:sol4}):

\begin{equation}
\small
\left\{
\begin{aligned}
0  < p_{000} & <  1/2 \\
0  < p_{001} & <  1/2(1 - 2 p_{000}) \\
0 < p_{010} & <  1/2 (1 - 2 p_{000} - 2 p_{001})  \\
p_{011} & =  1/2 (1 - 2 p_{000} - 2 p_{001} - 2 p_{010})  \\
p_{100} & =  p_{011}\\
p_{101} & =  1/2 (2 p_{010} + 2 p_{011} - 2 p_{100})\\
p_{110} & =  1/2 (2 p_{001} + 2 p_{011} - 2 p_{100})\\
p_{111} & =  p_{000} + p_{001} + p_{010} + p_{011} - p_{100} - p_{101} - p_{110}\\
\end{aligned}
\right. .
\label{eq:sol4}
\end{equation}

We now consider another example that is based on real data reported in Table~\ref{tab:fienberg-real}. 
Such dataset has been analyzed in \cite{fienberg:07} to illustrate interactions in three-way tables. 
In this example, thirty patients suffering from lymphocytic lymphoma who responded to a course of combination chemotherapy are classified by sex ($X_2$) and cell type ($X_3$). 
With a log-linear analysis, the author concludes that the cell type is related to both sex and outcome of the therapy ($X_1$). By using Mathematica, we could find the solution of the constrained problem with same odds ratio structure of the original data but uniform margins. The new table entries are showed in Table~\ref{tab:fienberg-real}. 

In these examples the difference between the original table and the transformed one is less evident because the three marginal distributions are more balanced than those in Table \ref{tab:id7}. We also notice that the solution derived with our method coincides with the one obtained by applying the standard Iterative Proportional Fitting Procedure \cite{barthelemy2018mipfp,rudas2018lectures}.

\begin{table}[h]
\centering
\begin{tabular}{@{}lccc@{}} 
\toprule 
&  & \multicolumn{2}{c}{Response ($X_1$)}\\\cmidrule{3-4} 
Clinic ($X_3$) & Treatment ($X_2$) & Success & Failure \\ \midrule 
1  & A & 18  & 12  \\ 
 & & (0.252) & (0.103) \\
& B & 12 & 8\\ 
 & & (0.103) & (0.042) \\
2 & A & 2 &  8\\ 
& & (0.042) & (0.103) \\
& B & 8 &  32\\
 & & (0.103) & (0.252) \\  \bottomrule 
\end{tabular}

\label{tab:Agrestisub2}
        \caption{Response vs Treatment and Clinic, from \cite{agresti:12}. Counts and solution to problem Eq.~(\ref{eq:condodd3}) (in brackets).}
\label{tab:Agresti2}
\end{table}

  \begin{table}[h]
  \centering
 \begin{tabular}{@{}lccc@{}} 
 \toprule 
 &  & \multicolumn{2}{c}{Outcome ($X_1$)}\\\cmidrule{3-4} 
 Cell type ($X_3$)& Sex ($X_2$)& No Response & Response \\ \midrule 
 Nodular  & Male & 1 & 4\\ 
  & & (0.024) & (0.133) \\
 & Female & 2 & 6\\ 
  & & (0.065) & (0.278) \\
 Diffuse & Male & 12 &  1\\
  & & (0.305) & ( 0.040) \\
 & Female & 3 &  1 \\
  & & (0.105) & (0.050) \\ \bottomrule 
 \end{tabular}
 
 \caption{Response of Lymphoma patients to Combination Chemotherapy: Distribution by Sex and Cell type, from \cite{fienberg:07}. Counts and solution to problem Eq.~(\ref{eq:condodd3}) (in brackets).} \label{tab:fienberg-real}
\end{table}

\section{Conclusions and discussion}
\label{sec:4}

In this work, we present possible extensions of \cite{geenens2020copula} to multi-way contingency tables. 
The primary goal of our investigation is to derive a unique solution to the problem of searching for a new table with preserved dependence structure and uniform margins.
As discussed in Sect.~\ref{sec:2} and Sect.~\ref{sec:3}, there are infinite tables with uniform margins and a fixed 3-dimensional odds ratio. 
However, we show that the uniqueness is guaranteed when also imposing constraints on the conditional odds-ratios. We notice that this is not the only possible choice to ensure uniqueness. 
For example, we could add constraints on the uniformity of the 2-dimensional sections of the multi-way table. Such a choice results in searching for the probabilities $p_{ijk} \geq 0, \; i,j,k=0,1$ that satisfy the system of equations reported in the following Eq~(\ref{eq:newsyst}):
\begin{equation}
\left\{
\begin{aligned}
\frac{p_{000} p_{011} p_{101} p_{110}}{p_{001} p_{010} p_{100} p_{111}} &=\omega_\star \\
p_{000} +p_{001} +p_{010} +p_{011} + p_{100} +p_{101} +p_{110} +p_{111} &=1\\
p_{000} +p_{001} +p_{010} +p_{011} - p_{100} - p_{101} - p_{110} - p_{111} &= 0\\
p_{000} +p_{001} -p_{010} -p_{011} + p_{100} + p_{101} - p_{110} - p_{111} &= 0\\
p_{000} - p_{001} +p_{010} -p_{011} + p_{100} - p_{101} + p_{110} - p_{111} &= 0\\
{p_{000} +p_{001} -p_{010} -p_{011}} &{=0}\\
{ p_{010} +p_{011} - p_{100} - p_{101}}&{=0}\\
{ p_{100} +p_{101} -p_{110} -p_{111}} &{=0}\\
{p_{000} -p_{001} +p_{010} -p_{011}} &{=0}\\
{p_{001} +p_{011} - p_{100} -p_{110}} &{=0}\\
{ p_{100} -p_{101} +p_{110} -p_{111}} &{=0}\\
{p_{000} -p_{001}  + p_{100} -p_{101}}  &{=0}\\
{p_{001} -p_{010} +p_{101} -p_{110}} &{=0}\\
{p_{010} -p_{011}  +p_{110} -p_{111}} &{=0}
\end{aligned}
\right. .
\label{eq:newsyst}
\end{equation}
Interestingly, the unique solution to this problem is the generalization of Eq~(\ref{eq:sol2}) given by the following entries of the three-way table
\[
\left\{
\begin{aligned}
p_{000}= p_{011}=p_{101}=p_{110}=\frac{\sqrt[4]{\omega_\star}}{4(1+\sqrt[4]{\omega_\star})} \\
p_{000}= p_{011}=p_{101}=p_{110}=\frac{1}{4(1+\sqrt[4]{\omega_\star})} 
\end{aligned}
\right. .
\]
Although the solution is unique, this approach results in a constrained dependence of $(X_1,X_2)$, $(X_1,X_3)$, and $(X_2,X_3)$ which is then modified from the original table structure.
Therefore, it is not in line with the idea of maintaining the dependence structure of the original table.
Unique solutions could also be obtained under other types of stochastic constraints, such as imposing exchangeability. 
We will investigate this aspect in a follow-up study. Additionally, the proposed theory is applicable when there are no zeros in the table and the (conditional) odds ratios can be computed. In future work, we plan to analyze the impact of potential zeros in the approach in a similar fashion as we did for structural zeros of two-way tables in \cite{Fontana2023}.

\subsubsection*{Acknowledgements}

Roberto Fontana and Fabio Rapallo are members of the GNAMPA-INdAM group.

%
%

\end{document}